\documentclass[review]{elsarticle}

\usepackage{lineno,hyperref}
\modulolinenumbers[5]

\usepackage{epsfig}
\usepackage{graphics}
\usepackage{latexsym}
\usepackage{amsmath}
\usepackage{amssymb}
\usepackage{rotating}
\usepackage{subfigure}
\usepackage{bm}
\usepackage{color}
\usepackage{blindtext}
\usepackage{hyperref}

\journal{Journal of Nuclear Physics A}

\begin{document}

\begin{frontmatter}

\title{Study of the $ s-\bar s $ asymmetry in the proton}

\author{Muhammad Goharipour}

\address{School of Particles and Accelerators, Institute for Research in Fundamental Sciences (IPM), P.O.Box 19568-36681, Tehran, Iran}
\ead{muhammad.goharipour@ipm.ir}

\begin{abstract}
The study of $ s-\bar s $ asymmetry is essential to better understand of
the structure of nucleon and also the perturbative and nonperturbative mechanisms for
sea quark generation. Actually, the nature and
dynamical origins of this asymmetry have always been an interesting subject to 
research both experimentally and theoretically. One of the most powerful models can lead 
to $ s-\bar s $ asymmetry is the meson-baryon model (MBM). In this work, using a
simplified configuration of this model suggested by Pumplin, we calculate the $ s-\bar s $ asymmetry for different values of cutoff parameter $\Lambda$, to study the dependence of model to this parameter and also to estimate the theoretical uncertainty imposed on the results due to its uncertainty. Then,
we study the evolution of distributions obtained both at next-to-leading order
(NLO) and next-to-next-to-leading order (NNLO) using different evolution schemes.
It is shown that the evolution of the intrinsic quark distributions from a low initial
scale, as suggested by Chang and Pang, is not a good choice at NNLO using variable flavor number scheme (VFNS).
\end{abstract}


\end{frontmatter}


\section{Introduction}\label{sec:one} 
It is well known now that the factorisation theorem of Quantum Chromodynamics (QCD)~\cite{Collins:1989gx,Brock:1993sz} 
can provide a powerful tool for calculating cross sections of high energy 
processes, by dividing them to perturbative and nonperturbative parts.
In this respect, the nonperturbative objects such as the parton distribution functions (PDFs)~\cite{Alekhin:2013nda,Accardi:2016qay,Ball:2014uwa,Harland-Lang:2014zoa,Jimenez-Delgado:2014twa,Dulat:2015mca,
Aleedaneshvar:2016dzb,Alekhin:2017kpj,Ball:2017nwa,MoosaviNejad:2016ebo,Khanpour:2016uxh}, polarized PDFs~\cite{
Jimenez-Delgado:2014xza,Sato:2016tuz,Shahri:2016uzl,Khanpour:2017cha,Ethier:2017zbq,Khanpour:2017fey,Salajegheh:2018hfs}, nuclear PDFs~\cite{Kovarik:2015cma,Khanpour:2016pph,Eskola:2016oht,Klasen:2017kwb,Wang:2016mzo}, and fragmentation functions (FFs)~\cite{Nejad:2015fdh,Leader:2015hna,deFlorian:2017lwf,Soleymaninia:2017xhc} play an essential role for testing QCD, describing the experimental data, and
searching New Physics. Among them, the PDFs have always been of particular importance.
Actually, more accurate PDFs are very essential for theory predictions and then for better understanding of the perturbative mechanism of QCD and the structure of the nucleon.
Although, recent developments in theory calculations and experimental measurements have improved
our knowledge of PDFs to a large extent, the situation is not very satisfying
for the case of flavor and quark-antiquark asymmetries.
 
It is proven that the perturbative regime of QCD
can lead to the $ s-\bar s $ asymmetry in the proton sea through the 
QCD evolution at next-to-next-to-leading order (NNLO) or at three loops~\cite{Catani:2004nc}.
However, it is significant only in regions of small momentum fraction $ x $ and 
also inconsiderable in magnitude, so that cannot describe the present 
experimental evidences for the $ s-\bar s $ asymmetry. In this way, the nature and
dynamical origins of $ s-\bar s $ asymmetry (as well as the $\bar d - \bar u$ flavor asymmetry~\citep{Salajegheh:2017iqp})
have always been an interesting subject to research both experimentally and theoretically
(for a review see Refs.~\cite{Kumano:1997cy,Chang:2014jba,Salajegheh:2015xoa} and references therein).
It is believed now that the $ s-\bar s $ asymmetry in the proton must has a 
nonperturbative origin. In this view, there are two kinds of sea quarks in the proton:
``extrinsic'' and ``intrinsic'' sea quarks which have major differences with each other.
The first ones are produced perturbatively through the splitting of the gluons into $ q\bar q $ pairs
and are dominant at small $ x $ regions, while the later ones are produced through the
nonperturbative fluctuations of the nucleon state to five-quark
states or meson plus baryon states and are dominant at large $ x $ regions.
In recent years, the intrinsic quarks have been a subject of study by 
many investigators~\cite{Salajegheh:2017iqp,Salajegheh:2015xoa,Chang:2011vx,Chang:2011du,Chang:2014lea,Beauchemin:2014rya,Lyonnet:2015dca,Brodsky:2015fna,Aleedaneshvar:2016rkm,Rostami:2016dqi,An:2017flb,Hou:2017khm,Bednyakov:2017vck}.

Although the existence of the intrinsic quarks in the proton sea, for the first time, was
suggested in the study of charm quark component by Brodsky, Hoyer, Peterson, 
and Sakai (BHPS)~\cite{Brodsky:1980pb}, the possible manifestations of nonperturbative 
effects for the $ s-\bar s $ asymmetry was first discussed by Signal and 
Thomas~\cite{Signal:1987gz}, applying the meson cloud model (MCM).
Moreover, Brodsky and Ma~\citep{Brodsky:1996hc} proposed a light-cone baryon-meson fluctuation
model to calculate the $ s-\bar s $ asymmetry in the proton and found a significantly different result in analogy to the result of MCM. In recent years, these original ideas have been followed in many papers~\cite{Holtmann:1996be,Christiansen:1998dz,Cao:1999da,Cao:2003ny,Ding:2004ht,Cao:2005dt,Traini:2013zqa,Vega:2015hti} to shed further light on the $ s-\bar s $ asymmetry in view of the MCM and light-cone baryon-meson fluctuation model. Another way to calculate the $ s-\bar s $ asymmetry in the nucleon is using the chiral quark model (CQM)~\citep{Weinberg:1978kz,Manohar:1983md}. Actually, this model has been many successes so far both for describing the flavor asymmetry and quark-antiquark asymmetry in the nucleon~\cite{Eichten:1991mt,Szczurek:1996tp,Wakamatsu:1997en,Cheng:1997tt,Burkardt:1991di,Melnitchouk:1999mv,Wakamatsu:2003wg,Ding:2004dv,Nematollahi:2012eu,Wakamatsu:2014asa,Wang:2016eoq,Wang:2016ndh}. In addition to these models, 
there is also a model called the scalar five-quark model 
suggested by Pumplin~\cite{Pumplin:2005yf} which can give us the intrinsic components of the quark sea. It is worth noting in this context that the BHPS and scalar five-quark models cannot give us any
asymmetry between the quark and antiquark distributions in the nucleon,
while the MCM, CQM and light-cone baryon-meson fluctuation
model can lead to this asymmetry.

Experimentally, the measurements
of charm production with dimuon events in the final state in deep inelastic scattering (DIS)~\cite{Dore:2011qe,Alberico:2001sd,Bazarko:1994tt,Mason:2007zz,KayisTopaksu:2011mx,Samoylov:2013xoa},
and also $ W $ boson production in association with a single charm quark in proton-proton 
collisions~\cite{Chatrchyan:2013uja,Aad:2014xca} can give us valuable information
about the $ s-\bar s $ asymmetry in the proton. It is believed that 
the anomaly seen by the NuTeV experiment~\cite{Zeller:2001hh} in the extraction of the Weinberg
angle from neutrino-nucleus DIS can be explained by assuming the $ s-\bar s $ asymmetry
in the proton sea. For example, in Refs.~\cite{Ding:2004ht} and~\cite{Ding:2004dv}, some first proposals to relate the $ s-\bar s $ asymmetry to the NuTeV anomaly with phenomenological
success have been presented according to the light-cone baryon-meson fluctuation model and CQM, respectively (for a review, see Ref.~\cite{Ma:2005yt}). In addition, there have been further phenomenological applications of the $ s-\bar s $ to some experimental facts~\cite{Gao:2008ch,Chi:2014xba,Du:2017nzy}. For example, in Ref.~\cite{Du:2017nzy}, the authors have indicated that the difference between $ \Lambda $ and $ \bar\Lambda $ production is related to the asymmetric strange-antistrange distribution
inside the nucleon.

Although the lowest moment of the $ s-\bar s $ asymmetry, 
$ \langle s-\bar s \rangle=\int_0^1[s(x)-\bar s(x)]dx $, is equal to zero
because there is no net valence strange quark in the nucleon, 
the second moment of this asymmetry, $ S^-\equiv \langle x(s-\bar s) \rangle=\int_0^1 dx~x[s(x)-\bar s(x)] $,
can have a non-zero value. Actually, there is great interest to determine $ S^- $ both experimentally and theoretically.
For example, we can refer to the next-to-leading order (NLO) analysis of dimuon
events from neutrino (antineutrino) DIS with the nucleon performed by the NuTeV Collaboration~\cite{Mason:2007zz}
which leaded to $ S^-=0.00196\pm 0.00046\pm 0.00045 $ at momentum transverse squared $ Q^2=16 $ GeV$ ^2 $ and also the analysis performed by Barone \textit{et al.}~\cite{Barone:1999yv} using 
a wide range of related experimental data which leaded 
to $ S^-=0.0020\pm 0.0005 $ at $ Q^2=20 $ GeV$ ^2 $. Moreover, in the first global analysis including the CCFR and NuTeV dimuon data~\cite{Goncharov:2001qe}, the authors found $ -0.001< S^- <0.004 $~\cite{Olness:2003wz}.
However, it should be noted that a value of $ S^-= 0.0\pm0.002 $ 
has been obtained by Bentz \textit{et al.}~\cite{Bentz:2009yy} at same scale 
that is consistent with no $ s-\bar s $ asymmetry.
As an example of theoretical estimation of $ S^-$, one can refer to Ref.~\cite{Vega:2015hti}
where the authors achieved various values for $ S^-$ from 0.00047 to 0.00157.

As mentioned, one of the most powerful intrinsic quark models can lead to both flavor
and quark-antiquark asymmetries in the nucleon is the meson-baryon model (MBM).
In the MBM framework, the nucleon sometimes
fluctuates to a virtual baryon plus a meson state ($ N\longrightarrow MB $).
Contributions to the strange sea can come, for example, from fluctuations
to the two-body state $ K^+\Lambda^0 $, where $ K^+ $ is a $ u\bar s $ meson and 
$ \Lambda^0 $ is a $ uds $ baryon. Although the MBM formalism is rather complicated computationally,
Pumplin~\cite{Pumplin:2005yf} has introduced a more simple configuration based on original concepts
of this model and used it, for the first time, for calculating the intrinsic charm in the nucleon. 
A similar study has also been performed in the 
case of intrinsic strange~\cite{Salajegheh:2015xoa}. It is worth noting here that in Pumplin model,
the quantity plays an important role is the cutoff parameter $\Lambda$ so that its chosen value can change the final results.
To be more precise, we can consider a theoretical uncertainty on the obtained distributions due to the 
$\Lambda$ variations. Another important issue in this respect, is the evolution of the intrinsic 
quark distributions using the DGLAP equations~\cite{Gribov:1972ri}. It is shown that using the non-singlet evolution equations, one can 
determine the intrinsic quark distributions at higher values of $ Q^2 $~\cite{Lyonnet:2015dca}.
In Refs~\cite{Chang:2011vx,Chang:2011du,Chang:2014lea}, Chang and Pang were also suggested that the evolution of the
intrinsic quark distributions from a very lower initial scale such as $ \mu_0=0.3 $
or 0.5 GeV leads to a better fit to the experimental data. In this work, focusing
on the $ s-\bar s $ asymmetry in the proton, we are
going to investigate with more precision about two issues: 1) the dependence of the Pumplin model to
the cutoff parameter $\Lambda$ and the amount of the theoretical uncertainty imposed on the results due to its variation, and 2) the evolution of $ x(s-\bar s) $ distribution and the validity of the Chang and Pang suggestion for different evolution
schemes and also the order of evolution. 

The paper is organised as follows. In Sec.~\ref{sec:two}, we
review briefly the original MBM formalism and also the simplified configuration of it suggested by Pumplin, and present the procedure for calculating the $ s-\bar s $ asymmetry in the proton.
In Sec.~\ref{sec:three}, we calculate this asymmetry using
different values of cutoff parameter $\Lambda$ to study the dependence of the model to this parameter and also 
to estimate the theoretical uncertainty caused by it.
The study of the evolution of $ s-\bar s $ asymmetry and then its behaviour at higher $ Q^2 $
is performed in Sec.~\ref{sec:four}. We evolve the distributions obtained both at
NLO and NNLO approximation using fixed flavor
number scheme (FFNS) and variable flavor number scheme (VFNS).
Finally, we summarize our results and conclusions in Sec.~\ref{sec:five}.

%
%
\section{Meson-baryon model framework}\label{sec:two}
As mentioned in the Introduction, the possible intrinsic contribution to the
$ s-\bar s $ asymmetry was pointed out for the first time by Signal and
Thomas~\cite{Signal:1987gz} applying the MCM. The first calculation of the $ s-\bar s $ asymmetry according the light-cone baryon-meson fluctuation model was performed by Brodsky and Ma~\cite{Brodsky:1996hc}.
These original works were followed in other 
papers~\cite{Holtmann:1996be,Christiansen:1998dz,Cao:1999da,Cao:2003ny,Ding:2004ht,Cao:2005dt,Traini:2013zqa,Vega:2015hti} in order to further investigation
in this subject. Note, such asymmetry that is a natural consequence of SU(3) symmetry breaking in QCD,
can also be achieved for the case of charm quark~\cite{Melnitchouk:1997ig,Paiva:1996dd,Hobbs:2013bia}.
The main virtue of the meson-baryon model compared with the BHPS and
scalar five-quark models is that it can lead to the $ s-\bar s $
asymmetry in the nucleon sea. To be more precise, there are two origins cause this 
asymmetry: First is the difference between the probability distributions of the meson and
baryon in the proton, and second is the difference between the strange and antistrange
distributions in the baryon and meson, respectively.

According to the MBM formalism, we can consider that the wave function of the nucleon
is a series involving bare nucleon and meson-baryon
states so that we can write it as~\cite{Holtmann:1996be}
\begin{align}
\vert N \rangle_{physical} &= \sqrt{Z}\vert N \rangle_{bare} 
+ \sum_{MB}\sum_{\lambda\lambda '} \int dy~d^2\mathbf{k}_\perp \phi_{MB}^{\lambda\lambda '}(y,k^2_\perp)\nonumber\\
&\times \vert M^\lambda(y,\mathbf{k}_\perp) ; B^{\lambda '}(1-y,-\mathbf{k}_\perp) \rangle.
\label{eq1}
\end{align}
In the above formula, the first term is related to the ``bare'' nucleon and $ Z $
is the wave function renormalization constant. Moreover, the probability amplitude 
of the Fock state containing  a virtual meson $ M $ with longitudinal momentum 
fraction $ y $, transverse momentum $ \mathbf{k}_\perp $, and helicity $ \lambda $, 
and a virtual baryon $ B $ with longitudinal momentum fraction $ 1-y $, 
transverse momentum $ -\mathbf{k}_\perp $, and helicity $ \lambda ' $ 
denoted by $ \phi_{MB}^{\lambda\lambda '}(y,k^2_\perp) $.
Since there are no interactions among the $ q $ and $ \bar q $ in the meson-baryon 
components during the interaction with the hard photon in the deep inelastic process,
the contributions to the quark and antiquark distributions of the nucleon
can be expressed as a convolution between the distribution functions of quarks or antiquarks in
the hyperon or meson with the fluctuation functions of these hadrons.
In the case of strange quark, for spin dependence distributions we have~\cite{Cao:2003ny}
\begin{align}
s^N(x)=\sum_{BM}\,
    \int_x^1 \frac{d\bar y}{\bar y}\,
	  f_{BM}(\bar y)\, s_B\Big(\frac{x}{\bar y}\Big)
	  \label{eq2} \\
	  \bar{s}^N(x)=\sum_{MB}\,
    \int_x^1 \frac{dy}{y}\,
	  f_{MB}(y)\, \bar{s}_M\Big(\frac{x}{y}\Big)
\label{eq3}
\end{align}
where $ {\bar y}\equiv1-y $. The fluctuation function $  f_{BM}(\bar y) $ 
describes the probability of a nucleon $ N $ fluctuating into a baryon $ B $ with longitudinal momentum fraction $ \bar y $,
while $ f_{MB}(y) $ is related to the nucleon's fluctuation into a meson $ M $ with longitudinal momentum fraction $ y $.
Meanwhile, the $ s_B $ and $ \bar{s}_M $ are the strange and antistrange 
distributions in the baryon and meson, respectively. According to the MBM, 
the fluctuation functions are related to the amplitudes $ \phi_{MB}^{\lambda\lambda '} $ as follows
\begin{equation}
f_{MB}(y)=\sum_{\lambda\lambda '}\int_0^\infty \, d \mathbf{k}^2_\perp \vert \phi_{MB}^{\lambda\lambda '}(y,k^2_\perp) \vert ^2.
\label{eq4}
\end{equation}
Note that we must also have the relation $ f_{BM}(\bar y)=f_{MB}(y) $ to guarantee the conservation of momentum and
charge. Using the effective meson-nucleon Lagrangians, we can drive
the meson-baryon probability amplitude $ \phi_{MB}^{\lambda\lambda '} $ as
a function of the nucleon, baryon and meson masses, the invariant mass squared of 
the meson-baryon Fock state and also the vertex functions which contain the spin dependence of
the amplitude~\cite{Holtmann:1996be}. As it stands, the MBM formalism
is rather complicated computationally.

Beside the above presented configuration for MBM, 
Pumplin~\cite{Pumplin:2005yf} has introduced another configuration based on original concepts
of this model that is simpler computationally. According to the Pumplin model,
we can use an overall relation to model both the meson-baryon probability distribution 
in the nucleon (equivalent to the fluctuation function) and the constituent quark distributions in
the baryon or meson. To be more precise, the light-cone probability distributions 
can be derived directly from Feynman diagram rules and written as~\cite{Pumplin:2005yf}
\begin{align}
dP&= \frac{g^2}{(16\pi^2)^{N-1}(N-2)!}\,
    \prod_{j=1}^N dx_j\, \delta\left( 1-\sum_{j=1}^N x_j\right)\nonumber\\
  &\quad\times\int_{s_0}^\infty ds\, \frac{(s-s_0)^{N-2}}{(s-m_0^2)^2}\,
    |F(s)|^2,
\label{eq5}
\end{align}
where
\begin{equation}
s_0 = \sum_{j=1}^N \frac{m_j^2}{x_j}\,,
\label{eq6}
\end{equation}
and $ N $ is the number of particles with masses 
$ m_1, m_2, ..., m_N $ and spin $ 0 $ which are coupled
to a point scalar particle with mass $ m_0 $ and spin $0$ by a point-coupling
$ ig $. In Eq.~\ref{eq5}, the form factor $F^2$ is a function of $ s $ and
has been included to consider further suppression of high-mass states
and characterize the dynamics of the bound state. Two exponential and power-law
forms have been suggested for $F^2$ as follows:
\begin{equation}
|F^2(s)| =\exp [-(s-m_0^2)/\Lambda^2],
\label{eq7}
\end{equation}
\begin{equation}
|F^2(s)|= (s +\Lambda^2 ) ^ {-n}.
\label{eq8}
\end{equation}
The cutoff parameter $\Lambda$ can take any value between 2 and 10 GeV.

Now, as suggested by Pumplin, having both the meson-baryon Fock state probability distribution 
in the nucleon and the constituent quark distributions in the baryon or meson, we can use the following
relation defined as convolutions of the distributions to calculate the intrinsic 
quark and antiquark distributions in the nucleon,
\begin{align}
{{dP} \over {dx}} &= 
\int_0^1 \! dx_1 \, f_1(x_1) \int_0^1 \! dx_2 \, f_2(x_2) \, 
\delta(x - x_1 x_2)\nonumber\\ &=
\int_x^1{{dy} \over {y}} \, f_1(y) \, f_2(x/y).
\label{eq9}
\end{align}
For the case of intrinsic strange, we can consider six fluctuations as follows:
\begin{eqnarray}
p&\longrightarrow & K^+(u\bar{s})\Lambda^0 (uds),\nonumber\\
p&\longrightarrow & K^0(d\bar{s})\Sigma^+ (uus),\nonumber\\
p&\longrightarrow & K^+(u\bar{s})\Sigma^0 (uds),\nonumber\\
p&\longrightarrow & K^{*+}(u\bar{s})\Lambda^0 (uds),\nonumber\\
p&\longrightarrow & K^{*0}(d\bar{s})\Sigma^+ (uus),\nonumber\\
p&\longrightarrow & K^{*+}(u\bar{s})\Sigma^0 (uds).
\label{eq10}
\end{eqnarray}
However, due to high equality of the $ K^0 $ and $ K^+ $, $ K^{*0} $ 
and $ K^{*+} $, and also $ \Lambda $ and $ \Sigma $ physical masses, 
only two states $ K^+\Lambda^0 $ and $ K^{*+}\Lambda^0 $ 
lead to the different shapes for $ s $ and $ \bar s $ distributions in the nucleon and thus 
the $ s-\bar s $ asymmetry~\cite{Salajegheh:2015xoa}. As can be seen, since
the involved physical masses of hadrons are determined with high accuracy from the experimental
informations, the main parameter that its value can change the final results in this simplified
configuration of the meson-baryon model is 
the cutoff parameter $\Lambda$. Actually, by calculating the distributions with different values of $\Lambda$,
we can estimate the theoretical uncertainty imposed on the results due to the 
$\Lambda$ variation. In the next section, we present the numerical results for the
$ s-\bar s $ asymmetry in the proton and study in details the dependence of the model to cutoff parameter $\Lambda$.

%
\section{The $ s-\bar s $ asymmetry in the proton}\label{sec:three} 
The accurate determination of PDFs 
in the nucleon has always been an important subject in high energy physics.  
Since the PDFs are the nonperturbative objects, they have to be constrained in a global
analysis to experimental data~\cite{Alekhin:2013nda,Accardi:2016qay,Ball:2014uwa,Harland-Lang:2014zoa,Jimenez-Delgado:2014twa,Dulat:2015mca,
Aleedaneshvar:2016dzb,Alekhin:2017kpj,Ball:2017nwa,MoosaviNejad:2016ebo,Khanpour:2016uxh}.
In this vein, the determination of the gluon distribution both in the 
nucleon~\cite{Goharipour:2017rjl} and nuclei~\cite{Goharipour:2017uic}, and also the sea quark distributions and possible asymmetries between them
is of particular importance. Nowadays, thanks to many experiments
provide a wide range of accurate data including the DIS, 
$ pp $ and $ p\bar p $ collider measurements, our knowledge of the valence quarks, 
and to a large extent, sea quarks and gluon distributions is satisfying. However, it is not enough
in the case of flavor and quark-antiquark asymmetries. 

After the observation of the Gottfried sum rule violation by the New Muon Collaboration (NMC)
in measuring the proton and deuteron $ F_2 $ structure functions~\cite{Amaudruz:1991at} from deep inelastic muon scattering on 
hydrogen and deuterium targets, it is believed that the 
$\bar u$ and $\bar d$ distributions in the nucleon are surely different. 
The antiquark flavor asymmetry $ \bar d-\bar u $ was then confirmed by
the HERMES Collaboration~\cite{Ackerstaff:1998sr}
in a semi-inclusive DIS (SIDIS) experiment and the FNAL E866/NuSea Collaboration~\cite{Hawker:1998ty} 
by measuring $ pp $ and $ pd $ Drell-Yan processes. All measurements 
demonstrated that there is a $ \bar d $ excess over $ \bar u $ in the nucleon sea.

\begin{figure}[t!]
\centering
\includegraphics[width=8.5cm]{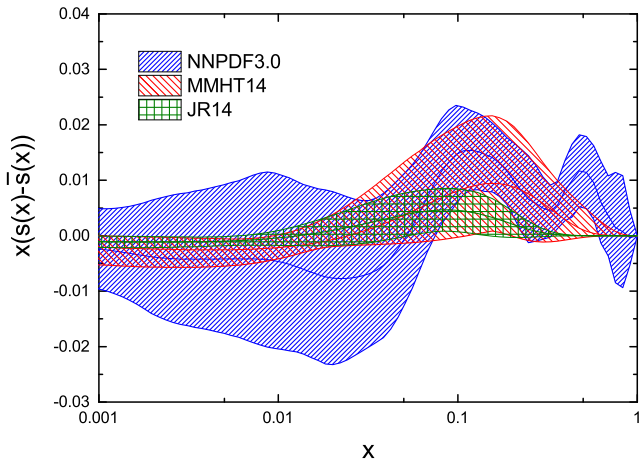}
\caption{The NNLO $ x(s(x)-\bar s(x)) $ distribution and its uncertainty
from NNPDF3.0~\cite{Ball:2014uwa} (blue band), MMHT14~\cite{Harland-Lang:2014zoa} (red band) and 
JR14~\cite{Jimenez-Delgado:2014twa} (green band) PDF sets at scale $ Q^2=4 $ GeV$ ^2 $. 
Results adopted from Ref.~\cite{GhasempourNesheli:2016pog}.}
\label{fig:fig1}
\end{figure}

Although there are some relatively clear evidences for the $ s-\bar s $ asymmetry of
the nucleon sea~\cite{Mason:2007zz,Zeller:2001hh,Barone:1999yv}, in the contrary, some
analyses are consistent with the strange-antistrange symmetry~\cite{Bentz:2009yy,Barone:2006xj}.
Since these results prevent a definitive conclusion on the existence of $ s-\bar s $ asymmetry in the nucleon,
various groups choose different approaches to deal with this asymmetry in
the global analysis of PDFs. For example, the NNPDF3.0~\cite{Ball:2014uwa}, MMHT14~\cite{Harland-Lang:2014zoa} 
and JR14~\cite{Jimenez-Delgado:2014twa} have considered $ s-\bar s $ asymmetry 
in their works while the ABM12~\cite{Alekhin:2013nda}, CT14~\cite{Dulat:2015mca} 
and CJ15~\cite{Accardi:2016qay} have assumed $ s(x)=\bar s(x) $. A
comparison of the results obtained for the $ x(s(x)-\bar s(x)) $ distribution and its uncertainty from
the NNLO analyses of NNPDF3.0, MMHT14 and JR14 at scale $ Q^2=4 $ GeV$ ^2 $
have been shown in Fig.~\ref{fig:fig1} by the blue, red and green shaded band, respectively.
As can be seen, the results have major differences can be related to the differences in used 
phenomenological approaches. For example, the $ x(s(x)-\bar s(x)) $ distribution 
from the NNPDF3.0, unlike two other PDF sets, has magnitude even at larger $ x $.
Furthermore, the NNPDF3.0 uncertainty is comparatively large at smaller $ x $.

As mentioned in the Introduction, beside the phenomenologically determination of
the $ s-\bar s $ asymmetry, it can be calculated
directly using some theoretical models based on the light-cone framework.
In the previous section, we introduced the MBM formalism and also Pumplin model for calculating the
intrinsic quark and antiquark distributions in the nucleon and the asymmetry
between them. In this section, we present the numerical results for the case of 
strange quark and study in details the dependence of the model to cutoff parameter $\Lambda$.
Its variation can be recognized as a source for generating the theoretical uncertainty imposed on the results. 
In this respect, we first calculate the $ x(s(x)-\bar s(x)) $ distributions related to 
the $ K^+\Lambda^0 $ and $ K^{*+}\Lambda^0 $ states separately and then
sum their results to obtain the total distributions.

For the case of $ K^+\Lambda^0 $ state, we calculate its probability distribution
in the proton using Eq.~\eqref{eq5} with $ N=2 $ and 
$ F^2\varpropto (s_{K\Lambda}+\Lambda^2_{p})^{-2}$. The same 
calculation is preformed using $ F^2\varpropto (s_{u\bar s}+\Lambda^2_{K})^{-2}$ to model the $ u\bar s $ 
distribution in $ K^+ $. For modeling the $ uds $ distribution in 
$ \Lambda^0 $ we use Eq.~\eqref{eq5} with $ N=3 $ and $ F^2\varpropto 
(s_{uds}+\Lambda^2_{\Lambda})^{-2}$. 
Since the quarks do not exist as free particles, their masses cannot be measured directly 
and then they are arbitrary parameters in QCD~\cite{Fritzsch:2014yra}. In this work, for up and down quark masses
we choose $ m_u=m_d=m_p/3 $ where $ m_p $ is the mass of proton, while
for the strange quark we should consider a larger value, for example, $ m_s=m_{\bar s}=0.5 $ GeV.
The physical messes of the proton, $ K^+$ meson and $\Lambda^0 $ baryon
are taken to be equal to 0.938, 0.4937 and 1.1157 in GeV, respectively. In this way,
the only quantities remain are the cutoff parameters $ \Lambda $. In Ref.~\cite{Pumplin:2005yf},
it has been suggested that we can choose any value between 2 and 10 for $ \Lambda_p $
and between 1 and 4 for $ \Lambda_K $ or $ \Lambda_\Lambda $. This can effect both the
shape and magnitude of the final results. Therefore, we have an uncertainty for the obtained distributions due to the 
$\Lambda$ uncertainty. Since it is interesting to study the dependence of the model to cutoff parameter $ \Lambda $
and also to estimate the theoretical uncertainty imposed on the results, we calculate distributions
for $ \Lambda_p=2, 4, 10 $ GeV and $ \Lambda_K=\Lambda_\Lambda=1, 2, 4 $ GeV. After the calculation of the 
$ K^+\Lambda^0 $ probability distribution in the proton, $ s $ distribution in $\Lambda^0$
and $ \bar s $ distribution in $ K^+$, we can calculate the corresponding 
$ s $ and $ \bar s $ distributions in the proton by doing the convolution of Eq.~\eqref{eq9}
and then the asymmetry between them. As a last point, note that we normalize
all distributions to 100\% probability so that the quark number 
condition $ \int_0^1 dxf(x)=1 $ is satisfied where $ f $ is the $ s $ or $ \bar s $ distribution
in related baryon or meson, respectively.

\begin{figure}[t!]
\centering
\vspace{0.5cm}
\includegraphics[width=8.0cm]{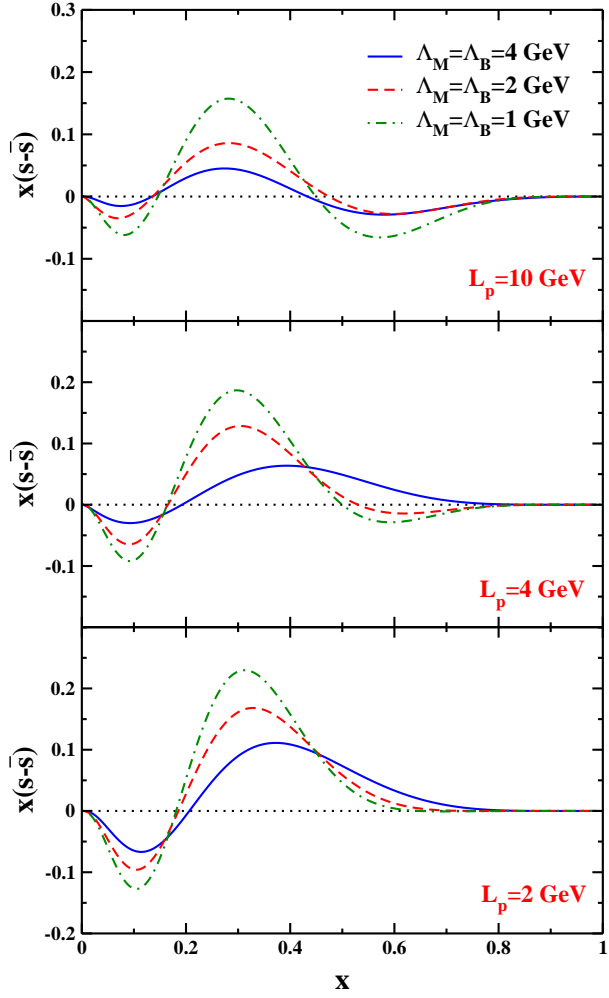}
\caption{The $ x(s(x)-\bar s(x)) $ distributions as a function of $ x $ obtained using the Pumplin model for $ K^+\Lambda^0 $ state
with $ \Lambda_p=10 $ (top panel), 4 (middle panel) and 2 (bottom panel).
In each panel, the blue solid, red dashed and green dotted-dashed curves are corresponding to the 
distributions for $ \Lambda_K=\Lambda_\Lambda=4 $, 2 and 1, respectively.}
\label{fig:fig2}
\end{figure} 
Fig.~\ref{fig:fig2} shows the results obtained for $ x(s(x)-\bar s(x)) $ distributions
as a function of momentum fraction $ x $ for $ K^+\Lambda^0 $ state. 
The top, middle and bottom panels are related to results for $ \Lambda_p=10 $, 4 and 2 GeV, respectively. 
In each panel, the blue solid, red dashed and green dotted-dashed curves are corresponding to the 
distributions for $ \Lambda_K=\Lambda_\Lambda=4 $, 2 and 1 GeV, respectively.
As can be seen from the top panel of Fig.~\ref{fig:fig2}, for $ \Lambda_p=10 $ there
are two negative areas in $ x(s(x)-\bar s(x)) $ distributions, one in smaller $ x $ and the other in larger $ x $,
and at medium $ x $ the asymmetry is clearly positive. Comparing the three panels, one can see that 
as $ \Lambda_p $ decreases, the negative area at larger $ x $ disappears and also the
magnitude of the distributions increases. Another conclusion one can draw from this figure is that 
for a fixed value of $ \Lambda_p $, the magnitude of the distributions is decreased when the value 
of $ \Lambda_K=\Lambda_\Lambda $ increases. In overall, we can conclude that the
$ x(s(x)-\bar s(x)) $ distribution resulted from $ K^+\Lambda^0 $ state, both in shape and
magnitude is very sensitive to the value of cutoff parameter $\Lambda$. 
\begin{figure}[t!]
\centering
\vspace{0.5cm}
\includegraphics[width=8.0cm]{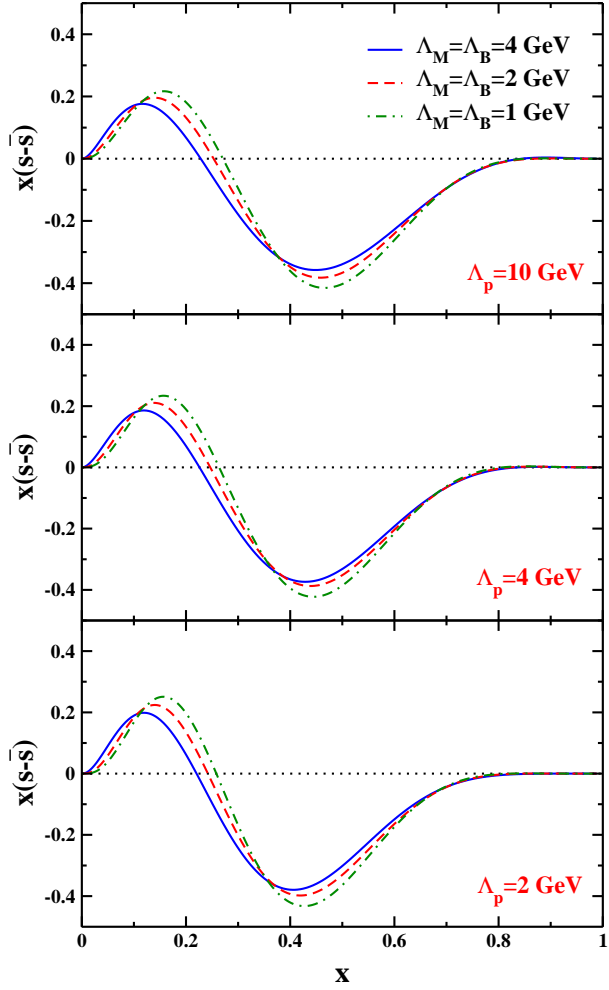}
\caption{Same as Fig.~\ref{fig:fig2} but for $ K^{*+}\Lambda^0 $ state.}
\label{fig:fig3}
\end{figure} 

As mentioned in the previous section, another meson-baryon state
can lead to different shape for $ s $ and $ \bar s $ distributions in the nucleon and thus 
the $ s-\bar s $ asymmetry is the $ K^{*+}\Lambda^0 $ state. The calculation procedure 
for this case is as before, but it should be noted that in order to avoid mass singularity we
consider an effective mass $ m_{\bar s}=0.7 $ GeV for the antistrange 
in $ K^{*+} $ as suggested in Ref.~\cite{Salajegheh:2015xoa}. In fact,
with this choice, the relation $ m_{\bar s}+m_u>m_{K^*} $ is satisfied.
Fig.~\ref{fig:fig3} shows same results as Fig.~\ref{fig:fig2} but for $ K^{*+}\Lambda^0 $ state.
As can be seen, in this case, there are two overall regions in $ x(s(x)-\bar s(x)) $ distributions
for all three values of $ \Lambda_p $:
a positive region in smaller $ x $ and a negative one in medium and larger $ x $.
As before, when $ \Lambda_p $ decreases, the magnitude of the distributions increases but
the changes are not too drastic. By focusing on each panel separately, we find that 
as $ \Lambda_K=\Lambda_\Lambda $ increases, the magnitude of the 
distribution somewhat decreases and also it shifts slightly toward smaller $ x $.
In overall, we can say that the $ s-\bar s $ asymmetry resulted from $ K^{*+}\Lambda^0 $ state
in not very sensitive to the value of cutoff parameter $\Lambda$.

\begin{figure}[t!]
\centering
\vspace{0.5cm}
\includegraphics[width=8.0cm]{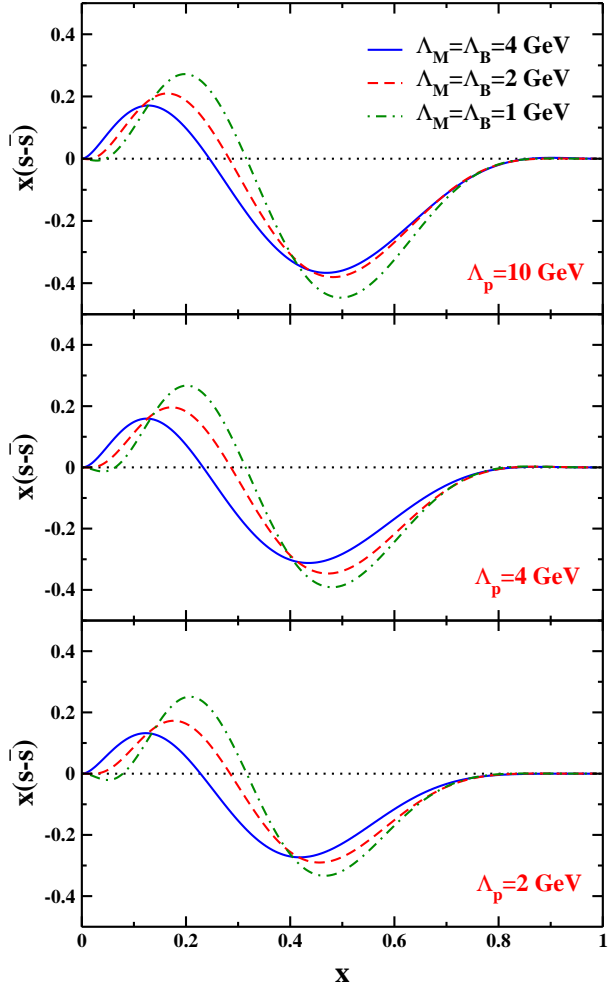}
\caption{The total $ x(s(x)-\bar s(x)) $ distributions in the proton from the Pumplin model as a function of $ x $
obtained by summing the results related to $ K^+\Lambda^0 $ and $ K^{*+}\Lambda^0 $ states 
with $ \Lambda_p=10 $ (top panel), 4 (middle panel) and 2 (bottom panel).
In each panel, the blue solid, red dashed and green dotted-dashed curves are corresponding to the 
distributions for $ \Lambda_K=\Lambda_\Lambda=4 $, 2 and 1, respectively.}
\label{fig:fig4}
\end{figure}

Now we can sum the results of $ x(s(x)-\bar s(x)) $ distributions for
$ K^+\Lambda^0 $ and $ K^{*+}\Lambda^0 $ states presented in Figs.~\ref{fig:fig2} and~\ref{fig:fig3}
to get the total results of the $ s-\bar s $ asymmetry in the proton.
The related results have been shown in Fig.~\ref{fig:fig4}. As one can see, 
just similar to the results of $ K^{*+}\Lambda^0 $ state, there is a positive and a negative
region at smaller and larger $ x $ in $ x(s(x)-\bar s(x)) $ distributions
for all three values of $ \Lambda_p $. However, there is an interesting finding 
in the total results which is in contrary with the results for $ K^+\Lambda^0 $ and $ K^{*+}\Lambda^0 $ states.
To be more precise, unlike before, the total distributions decrease in magnitude as
$ \Lambda_p $ decreases. Nevertheless, for a fixed value of $ \Lambda_p $, 
as $ \Lambda_K=\Lambda_\Lambda $ increases, the magnitude of the 
distribution somewhat decreases and also it shifts slightly toward smaller $ x $ just
like to the case of $ K^{*+}\Lambda^0 $. As a last point, note that the behavior of theoretical result obtained using the MBM is compatible with the phenomenological results of Fig.~\ref{fig:fig1}, only if we consider the result of $ K^+\Lambda^0 $ state.

\begin{figure}[t!]
\centering
\vspace{0.5cm}
\includegraphics[width=8.2cm]{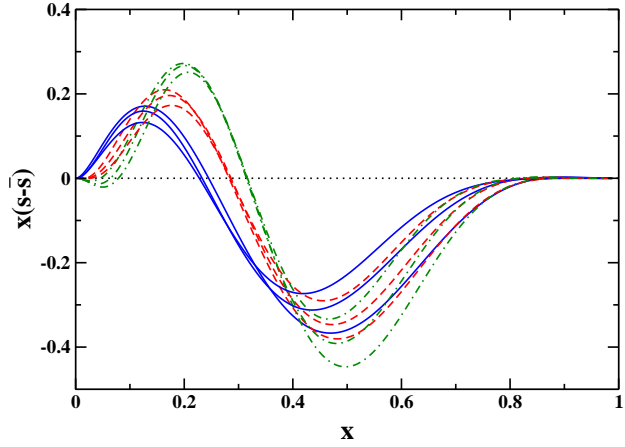}
\caption{Simultaneous plot of all total $ x(s(x)-\bar s(x)) $ distributions presented in Fig.~\ref{fig:fig4} which are 
corresponding to the different values of cutoff parameters $\Lambda$.}
\label{fig:fig5}
\end{figure}

\begin{figure}[t!]
\centering
\vspace{0.5cm}
\includegraphics[width=11.0cm]{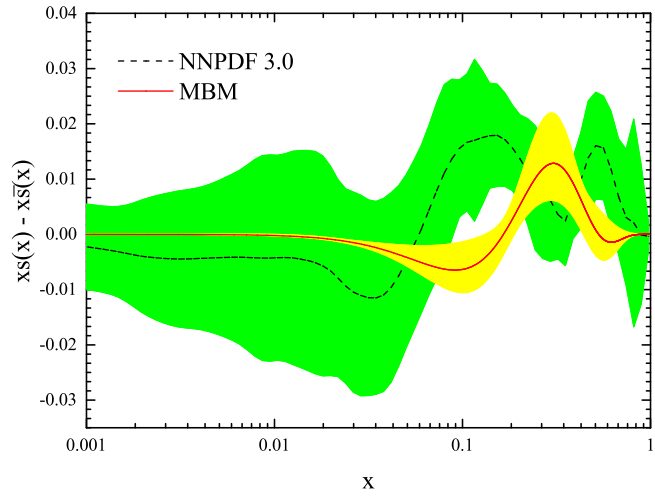}
\caption{A comparison between the theoretical result of MBM and phenomenologically obtained result of NNPDF3.0~\cite{Ball:2014uwa} for $ s-\bar s $ asymmetry. See text for further details.}
\label{fig:fig5b}
\end{figure}

We can discuss now about the theoretical uncertainty of the
$ s-\bar s $ asymmetry in the proton due to the uncertainties of the cutoff parameters $\Lambda$.
For this purpose, in Fig.~\ref{fig:fig5}, we have plotted simultaneously all distributions presented in three panels of Fig.~\ref{fig:fig4}
corresponding to different values of cutoff parameters $\Lambda$.
In this way, we can have an estimation of theoretical uncertainty in the $ s-\bar s $ asymmetry.
Note that the description of the curves is same as before. As a result, we can conclude that
the theoretical uncertainty of the $ s-\bar s $ asymmetry due to the variation of cutoff parameters $\Lambda$
is comparatively large in all regions of momentum fraction $ x $. However, for making an exact comparison between the results obtained in this section and phenomenological results for $ s-\bar s $ asymmetry, we can consider the distribution related to $ \Lambda_K=\Lambda_\Lambda=2 $ GeV and $ \Lambda_p=4 $ GeV as a central distribution and calculate an error band for it due to $\Lambda$ variations. The result has been shown in Fig.~\ref{fig:fig5b} (red solid curve with yellow band) and compared to the NNPDF3.0~\cite{Ball:2014uwa} result (black dashed curve with green band) at $ Q^2=1 $ GeV$ ^2 $. Note that the MBM result is corresponding to
a probability of 10\% for the $ K^+\Lambda^0 $ state as considered in Ref.~\cite{Brodsky:1996hc}.
As can be seen, there is a satisfying agreement between the theoretical prediction of MBM and the phenomenologically obtained result of NNPDF3.0 for $ s-\bar s $ asymmetry.
%
%
\section{The evolution of the $ s-\bar s $ asymmetry}\label{sec:four}
Having the total $ x(s(x)-\bar s(x)) $ distribution in the proton obtained
in the previous section, we are now ready to study its evolution and then the behaviour of
this asymmetry at higher $ Q^2 $. It is well known that the evolution of
PDFs is governed by the DGLAP integro-differential equations~\cite{Gribov:1972ri}.
Actually, if we have the parton densities as functions of $ x $ 
at an initial scale $ \mu_0^2 $, we can
obtain them at any arbitrary scale $ Q^2 $ by solving the DGLAP equations. Overall, these
equations can be divided into two general parts: singlet and non-singlet equations. 
A unique feature of the non-singlet equations is that the evolution of
a non-singlet distribution is independent of other patron densities and 
can be carried out solely. In this way, since the $ x(s(x)-\bar s(x)) $ is a non-singlet distribution,
we can evolve it to an arbitrary scale $ Q^2 $ no need to other PDFs.
The evolution can be performed by the {\tt QCDNUM} package~\cite{Botje:2010ay} both in fixed flavor
number scheme and variable flavor number scheme.
For our case (the evolution of the non-singlet distribution $ x(s-\bar s) $), the only deference
between these two schemes is in their procedure to deal with the number of active 
flavors $ n_f $ in the evolution of the strong coupling constant $ \alpha_s $.
To be more precise, in FFNS, $ n_f $ is kept fixed throughout the evolution, while
in VFNS, the flavor thresholds $ \mu_{c,b,t}^2 $ related to charm, bottom and top quark
masses are introduced and the value of
number of active flavors is changed from $ n_f $ to $ n_f +1$ at these thresholds 
(note that the number of active flavors is set to $ n_f=3 $ below the charm threshold).

In other to study the evolution of $ x(s(x)-\bar s(x)) $ in the proton,
we select the distribution obtained with $ \Lambda_p=4 $ and $ \Lambda_K=\Lambda_\Lambda=2 $
(see the previous section) and evolve it to $ Q^2=16 $ GeV$ ^2 $ as an example. The evolution
is preformed from two values of initial scale $ \mu_0=0.3 $ and 0.5 GeV 
as suggested by Chang and Pang in 
Refs.~\cite{Chang:2011vx,Chang:2011du,Chang:2014lea} and at both  NLO
and NNLO. Moreover, we present the results for both FFNS and VFNS to study the effect of 
chosen evolution scheme on the behaviour and also magnitude of evolved $ x(s(x)-\bar s(x)) $ distribution. 
It is worth noting that in all calculations, the value of strong coupling constant
at $ Z $ boson mass scale ($ \alpha_s(M_Z) $) is taken to be 0.118. 

Fig.~\ref{fig:fig6} shows the results for $ x(s(x)-\bar s(x)) $ at $ Q^2=16 $ GeV$ ^2 $ 
using FFNS with $ n_f=5 $. The black solid, red dashed, green long-dashed,
and blue dotted-dashed curves
are related to NLO with $ \mu_0=0.3 $ GeV, NLO with $ \mu_0=0.5 $ GeV, NNLO with
$ \mu_0=0.3 $ GeV and NNLO with $ \mu_0=0.5 $ GeV, respectively. As a first
conclusion, by comparing Fig.~\ref{fig:fig4} (see red dashed curve in middle panel) 
and~\ref{fig:fig6}, we can see that the distributions have been decreased in magnitude and also shifted to the smaller $ x $ 
due to the evolution. Another conclusion is that the results related to $ \mu_0=0.5 $ GeV have a larger
magnitude than $ \mu_0=0.3 $ GeV. Meanwhile, note that there is no considerable deference between the 
NLO and NNLO results in FFNS. 
\begin{figure}[t!]
\centering
\includegraphics[width=8.2cm]{fig6.eps}
\caption{The $ x(s(x)-\bar s(x)) $ distribution in the proton obtained using the Pumplin model and evolved from $ \mu_0=0.3 $ and 0.5 GeV
to $ Q^2=16 $ GeV$ ^2 $ at NLO and NNLO approximation using FFNS with $ n_f=5 $.}
\label{fig:fig6}
\end{figure} 
\begin{figure}[t!]
\vspace{0.5cm}
\centering
\includegraphics[width=8.2cm]{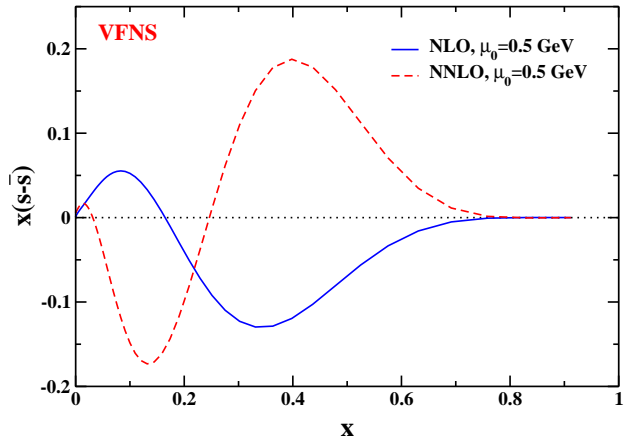}
\caption{The $ x(s(x)-\bar s(x)) $ distribution in the proton obtained using the Pumplin model and evolved from $ \mu_0=0.5 $ GeV
to $ Q^2=16 $ GeV$ ^2 $ at NLO and NNLO approximation using VFNS.}
\label{fig:fig7}
\end{figure} 

The results for $ x(s(x)-\bar s(x)) $ at $ Q^2=16 $ GeV$ ^2 $ 
using VFNS have been shown in Fig.~\ref{fig:fig7}. As can be seen, this figure includes only
the NLO (blue solid curve) and NNLO (red dashed curve) distributions for $ \mu_0=0.5 $ GeV.
Actually, since the value of $ \alpha_s $ becomes very large at both NLO and NNLO evolution
using VFNS with $ \mu_0=0.3 $ GeV, one gets a runtime error in {\tt QCDNUM} package preventing
the program to continue computation.
By comparing Figs.~\ref{fig:fig6}
and~\ref{fig:fig7}, we can see that, at NLO, the $ x(s(x)-\bar s(x)) $ distribution evolved using VFNS
behaves as one evolved using FFNS, but with a smaller magnitude in all range of $ x $.
A very surprising point can be raised from Fig.~\ref{fig:fig7} is that, at NNLO, 
the $ x(s(x)-\bar s(x)) $ distribution evolved using VFNS behaves 
quite different compared with one evolved using FFNS (and also with one evolved using VFNS at NLO). In fact, in this case, the position of positive and negative regions in 
$ x(s(x)-\bar s(x)) $ distribution has been exchanged due to the evolution. This finding
suggests that something is wrong, so that the result can be considered unphysical.
In other to further investigation on this issue, a good idea is using
another package for evolving the $ x(s(x)-\bar s(x)) $ distribution such as the {\tt PEGASUS}~\cite{Vogt:2004ns}.

\begin{figure}[t!]
\centering
\includegraphics[width=8.2cm]{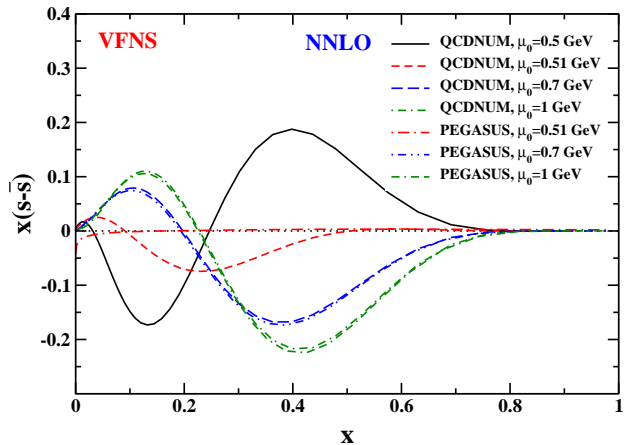}
\caption{The $ x(s(x)-\bar s(x)) $ distribution in the proton obtained using the Pumplin model and evolved from different initial scales $ \mu_0 $
to $ Q^2=16 $ GeV$ ^2 $ at NNLO approximation using VFNS by two packages {\tt QCDNUM}~\cite{Botje:2010ay} and {\tt PEGASUS}~\cite{Vogt:2004ns}.}
\label{fig:fig8}
\end{figure}

Fig~\ref{fig:fig8} shows a comparison between the {\tt QCDNUM} and {\tt PEGASUS} results for the evolution of
the $ x(s(x)-\bar s(x)) $ distribution from different initial scales $ \mu_0 $ 
to $ Q^2=16 $ GeV$ ^2 $ using the VFNS at NNLO. As can be seen, for the case of {\tt QCDNUM},
if one chooses a value grater than 0.5 GeV for initial scale $ \mu_0 $, for example
$ \mu_0= $ 0.51 (red short-dashed curve), 0.7 (blue long-dashed curve) 
and 1 GeV (green dotted-dashed curve), the result of evolution
will be natural too using VFNS at NNLO. However, when we choose exactly $ \mu_0=0.5 $ GeV, the result is changed dramatically. For a smaller value than 0.5 GeV (even $ \mu_0=0.49 $ GeV), 
one gets the runtime error due to excessive increase in the value of $ \alpha_s $.
For the case of {\tt PEGASUS}, the situation is a bit different. Actually, 
for $ \mu_0=$ 0.7 and 1 GeV, the {\tt QCDNUM} and {\tt PEGASUS} have same results.
But, for $ \mu_0=$ 0.51 GeV, their result is absolutely different.
To be more precise, in this scale, the result of {\tt QCDNUM} 
seems still natural, but it is clear that the result of {\tt PEGASUS} is unphysical.
It should be noted that the result of {\tt PEGASUS} for exactly $ \mu_0=$ 0.5 GeV has not been shown in Fig~\ref{fig:fig8}. In fact, in that scale, the value of $ \alpha_s $ in the {\tt PEGASUS} calculations
becomes infinity, so the program returns a ``NaN" value for the $ x(s-\bar s) $
in all $ x $. This dramatical behaviour of $ x(s-\bar s) $ distribution
under the evolution using the VFNS at NNLO from a very lower initial scale, can
be attributed to the excessive increase in the value of $ \alpha_s $, or maybe has another
reason should be carefully investigated in the future researches. Anyhow, the conclusion we can take with certainty from the results obtained in this section is that the choice of initial
scale $ \mu_0 $ is a very important ingredient in this respect, and the evolution of the intrinsic quark distributions from a low initial
scale, as suggested by Chang and Pang~\cite{Chang:2011vx,Chang:2011du,Chang:2014lea}, is not a good choice at NNLO using VFNS.
%

%
\section{Summary and conclusions}\label{sec:five}
One of the most powerful models can lead 
to $ s-\bar s $ asymmetry in the nucleon is the meson-baryon model (MBM).
According to the MBM formalism, we can consider that the wave function of the nucleon
is a series involving bare nucleon and meson-baryon states.
It can be shown that, among the possible states,
only two states $ K^+\Lambda^0 $ and $ K^{*+}\Lambda^0 $ 
lead to the different shapes for $ s $ and $ \bar s $ distributions in the nucleon and thus 
the $ s-\bar s $ asymmetry~\cite{Salajegheh:2015xoa}.
Although the MBM formalism is rather complicated computationally,
Pumplin~\cite{Pumplin:2005yf} has introduced a more simple configuration based on original concepts
of this model and used it, for the first time, for the calculation of the intrinsic charm in the nucleon. In Pumplin model,
the quantity plays an important role is the cutoff parameter $\Lambda$, so that its chosen value can change the final results.
In this way, we can consider a theoretical uncertainty on the distributions due to the 
$\Lambda$ variation. In this work, we calculated the $ s-\bar s $ asymmetry for
different values of $\Lambda$ to study the dependence of the model to this parameter and also 
to estimate the theoretical uncertainty imposed on the results due to its uncertainty.
As a result, we found that the $ x(s(x)-\bar s(x)) $ distribution resulted from $ K^+\Lambda^0 $ state, both in shape and
magnitude is very sensitive to value of $\Lambda$, while the related result
from $ K^{*+}\Lambda^0 $ state is not very sensitive to it. Then, we calculated
the total $ x(s(x)-\bar s(x)) $ distribution in the proton by summing the results obtained
for $ K^+\Lambda^0 $ and $ K^{*+}\Lambda^0 $ states. We concluded that
they decrease in magnitude as $ \Lambda_p $ decreases. Moreover, for a fixed value of $ \Lambda_p $, 
as $ \Lambda_K=\Lambda_\Lambda $ increases, the magnitude of the 
total distribution somewhat decreases and shifts slightly toward smaller $ x $. By comparing
all distributions obtained simultaneously, we found that
the theoretical uncertainty of the $ s-\bar s $ asymmetry due to the variation of the cutoff parameters $\Lambda$
is comparatively large in all regions of momentum fraction $ x $.
However, by calculating exactly the uncertainty of $ x(s(x)-\bar s(x)) $ distribution, we showed that there is a satisfying agreement between the theoretically prediction of MBM and the phenomenologically obtained result of NNPDF3.0, if one considers only the $ K^+\Lambda^0 $ state.  
We also studied the evolution of $ x(s(x)-\bar s(x)) $ distribution both at NLO 
and NNLO using different evolution schemes. The evolution
preformed from different initial scales $ \mu_0 $. As a result,
we found that the distributions are decreased in magnitude and also shifted to the smaller $ x $ 
due to the evolution using FFNS. Furthermore, the results related to $ \mu_0=0.5 $ GeV have a larger
magnitude than $ \mu_0=0.3 $ GeV and there is no considerable deference between the 
NLO and NNLO results in FFNS. By comparing the results of FFNS and VFNS, 
it was found that the evolved $ x(s(x)-\bar s(x)) $ distribution using VFNS at NLO
behaves as one evolved using FFNS, but with a smaller magnitude in all range of $ x $. 
Nevertheless, by performing the evolution of the $ x(s(x)-\bar s(x)) $ distribution using VFNS
at NNLO through two packages {\tt QCDNUM} and {\tt PEGASUS} and comparing their results,
we concluded that the choice of initial
scale $ \mu_0 $ is a very important ingredient in this respect. To be more precise, at NNLO and using VFNS with $ \mu_0= $ 0.5 GeV, the result of {\tt QCDNUM} 
behaves quite different compared with one evolved using FFNS (and 
also with one evolved using VFNS at NLO),
and using the {\tt PEGASUS}, one gets a ``NaN" value for $ x(s-\bar s) $ distribution
in all $ x $. This dramatical behaviour can
be attributed to the excessive increase in the value of $ \alpha_s $, or maybe has another
reason should be carefully investigated in the future researches. However, 
the conclusion we can take with certainty from the results obtained is that
the evolution of the intrinsic quark distributions from a low initial
scale, as suggested by Chang and Pang~\cite{Chang:2011vx,Chang:2011du,Chang:2014lea}, is not a good choice at NNLO using VFNS.

%
\section*{Acknowledgments}
I thank Michiel Botje and Hamzeh Khanpour for useful discussions and comments. I also thank School of Particles and Accelerators, Institute for Research in Fundamental Sciences (IPM) for financial support of this project.
%

{}

\end{document}